\documentclass[10pt,twocolumn,english,aps,pra,nofootinbib,superscriptaddress,showpacs,showkeys,longbibliography]{revtex4-1}
\usepackage[T1]{fontenc}
\usepackage{color}
\usepackage{babel}
\usepackage{multirow}
\usepackage{bbold}
\usepackage{url}
\usepackage{textcomp}
\usepackage{amssymb}
\usepackage{amsmath}
\usepackage{mathtools}
\usepackage{siunitx}
\usepackage{verbatim}
\usepackage{graphicx}
\usepackage{subfigure}
\graphicspath{ {images/} }
\usepackage{bbm, dsfont}
\usepackage{amsfonts}
\usepackage{babel}
\usepackage{mathptmx}

\definecolor{myurlcolor}{rgb}{0,0,0.7}
\definecolor{myrefcolor}{rgb}{0.8,0,0}
\usepackage[unicode=true,pdfusetitle, bookmarks=false,bookmarksnumbered=false,
bookmarksopen=false, breaklinks=false,pdfborder={0 0 0},backref=false,
colorlinks=true, linkcolor=myrefcolor,citecolor=myurlcolor,urlcolor=myurlcolor]
{hyperref}

\usepackage{hyperref}

\begin{document}

\title{Collective defense of honeybee colonies: experimental results and theoretical modeling}
\date{\today}
\author{Andrea L{\'o}pez-Incera}
\thanks{These two authors contributed equally.}
\affiliation{Institute for Theoretical Physics, University of Innsbruck, A-6020 Innsbruck, Austria}
\author{Morgane Nouvian}
\thanks{These two authors contributed equally.}
\affiliation{Department of Biology, Universit\"at Konstanz, 78457 Konstanz, Germany}
\affiliation{Zukunftskolleg, Universit\"at Konstanz, 78457 Konstanz, Germany}
\affiliation{Centre for the Advanced Study of Collective Behavior, Universit\"at Konstanz, 78457 Konstanz, Germany}
\author{Katja Ried}
\affiliation{Institute for Theoretical Physics, University of Innsbruck, A-6020 Innsbruck, Austria}
\author{Thomas M\"uller}
\affiliation{Fachbereich Philosophie, Universit\"at Konstanz, Fach 17, 78457 Konstanz, Germany}
\author{Hans J. Briegel}
\affiliation{Institute for Theoretical Physics, University of Innsbruck, A-6020 Innsbruck, Austria}
\affiliation{Fachbereich Philosophie, Universit\"at Konstanz, Fach 17, 78457 Konstanz, Germany}

\begin{abstract}
Social insect colonies routinely face large vertebrate predators, against which they need to mount a collective defense. To do so, honeybees use an alarm pheromone that recruits nearby bees into mass stinging of the perceived threat. This alarm pheromone is carried directly on the stinger, hence its concentration builds up during the course of the attack. Here, we investigate how individual bees react to different alarm pheromone concentrations, and how this evolved response-pattern leads to better coordination at the group level. We first present an individual dose-response curve to the alarm pheromone, obtained experimentally. Second, we apply Projective Simulation to model each bee as an artificial learning agent that relies on the pheromone concentration to decide whether to sting or not. If the emergent collective performance benefits the colony, the individual reactions that led to it are enhanced via reinforcement learning, thus emulating natural selection. Predators are modeled in a realistic way so that the effect of factors such as their resistance, their killing rate or their frequency of attacks can be studied. We are able to reproduce the experimentally measured response-pattern of real bees, and to identify the main selection pressures that shaped it. Finally, we apply the model to a case study: by tuning the parameters to represent the environmental conditions of European or African bees, we can predict the difference in aggressiveness observed between these two subspecies.
\end{abstract}
\maketitle

\section{Introduction}
Whoever has delighted in honey knows how much of a valuable food source a honeybee colony can be. To fend off the many predators attracted by this nutrient trove, honeybees have evolved stingers and a powerful venom efficient against vertebrates and invertebrates alike. But arguably their most important weapon is number: honeybees build a collective defense against intruders, stinging, threatening and harassing them in dozens or hundreds. Central to this response is the alarm pheromone carried directly on their stingers, whose banana-like scent is well known to beekeepers. When released, the sting alarm pheromone (SAP) alerts and attracts other bees, recruiting them to the site of the disturbance and priming them to sting. It is a chemically complex blend of over 40 compounds, but its main component, isoamyl acetate (IAA), is sufficient to trigger most of the behavioural response. The SAP has been extensively studied, both from the releaser end (production, dispersal) and from the recipient end (detection, reaction, role of the different compounds, role of the context in which it is perceived – reviewed in \citep{Nouvian16}). Despite this wealth of knowledge on the SAP, two important aspects of its function remain elusive: its quantitative action and the evolution of this response. In this study, we use a combination of \textit{in vivo} and \textit{in silico} methods to better understand how honeybees react to different concentrations of alarm pheromone, how this impacts the organization of the collective response, and which selection pressures might have driven the evolution of this defensive strategy. 

If they detect a threat, guard bees can disperse the SAP actively by raising their abdomen, extruding their stinger and fanning their wings. Alternatively, since the SAP is carried on the stinger itself, it is automatically released upon stinging. Thus, the SAP potentially carries information about the presence and location of a threat, but also about the magnitude of the attack already mounted against it. Several studies already demonstrated that the intensity of the response is correlated with the amount of alarm pheromone in the atmosphere \citep{Ghent62, Southwick85, Collins87, Lensky95}. While these studies provided valuable information, they all tested bees in groups - and often in field conditions, hence they could not establish an individual dose-response curve. Furthermore, in many cases the behavioral readout was rather indirect (e.g. attraction or fanning), only one study \citep{Lensky95} actually measured stinging frequency. To comprehend how each bee makes the decision to sting, and thus how the colony as a whole coordinates actions during a defensive event, an individually resolved dose-response curve is necessary. Here we took advantage of an assay developed a few years ago \citep{Nouvian15}, that measures stinging responses under controlled conditions, to fill this knowledge gap. We found that, indeed, there is a steep ramp up phase at low to medium alarm pheromone concentrations, in which the stinging likelihood of a bee increases together with the concentration. In addition, we show for the first time the existence of a second, decreasing phase at high pheromone concentrations. This is consistent with an anecdotal report that a high dose of IAA became repellent to bees \citep{Boch70}.

How to interpret this experimental curve? More precisely, what could have driven the evolution of such a response pattern? To address these questions, we resort to Projective Simulation \citep{BriegelCuevas12}, which is a simple model of agency that integrates a notion of episodic memory with a reinforcement learning mechanism. Projective Simulation (PS) allows for a realistic encoding of the sensory apparatus and motor abilities of the agents, which can perceive, make decisions and act as individuals. When interacting with other agents, these individual actions may influence the perceptions and responses of the rest of the ensemble, which in turn leads to the emergence of collective behavior. Crucially, neither the individual responses nor the interaction rules among agents are fixed in advance. Instead, they are developed throughout a learning process in which the agents' decisions are reinforced if they are beneficial under certain evolutionary pressures. All of these features make Projective Simulation a suitable framework to model behavioral experiments like the one presented above, since it allows us to analyze the observed responses from both the individual and the collective perspectives. Furthermore, we are able to draw conclusions about the possible evolutionary pressures that may have led to such behavior by means of the reinforcement learning process.

In this work, we model each bee as a learning PS agent and the colony as an ensemble of agents that undergoes repeated encounters with predators. During each encounter, bees can die from stinging but this also participates in deterring the predator, or they can be directly killed by the predator. Since the success of a bee colony is highly dependent on its available workforce, the overall performance of the colony is then evaluated by counting the number of bees that are still alive at the end of the encounter, and the individual response-pattern is rewarded accordingly. Hence, the whole simulation is similar to an evolutionary process, in which the behaviour of each generation of bees is passed on depending on its reproductive success. By investigating systematically the parameters of a simple but realistic predator model, we found that the initial ramp-up in stinging responsiveness was mainly driven by uncertainty on the predator detection (frequency of false alarms), and that the pheromone concentration at peak aggressiveness was dependent on the most resistant predator encountered. The second, decreasing phase could be interpreted as the combination of a self-limiting mechanism to avoid over-stinging and of a return to baseline due to lack of experience in this range of concentrations.

The native range of the Western honeybee (\textit{Apis mellifera}) spans a large part of Europe, Africa and Middle East Asia \citep{Requier19}, and thus includes a wide diversity of ecosystems. As a consequence, multiple subspecies exist that have adapted to local conditions. In particular, the African honeybees are known for having stronger defensive reactions than their European counterparts: they recruit more bees, do so more quickly, and are more persistent \citep{Breed04, Guzman04}. Part of the explanation resides in their higher sensitivity to the SAP \citep{Collins87}. As a final test of our model, we tune the parameters to represent the constraints that were hypothesized to drive the evolution of this striking difference in behaviour (mainly a higher predation rate in Africa). We show that, with this input, our model accurately predicts the different strategies adopted by each subspecies. Thus, this novel application of Projective Simulation to a group of agents with a common goal is promising for the study of social insects in general, and of the honeybee defensive behaviour in particular.

\section{Methods and model}
\subsection{Experimental material and methods}\label{SUBSEC exper. methods}
\subsubsection{Honeybees}

The experiments were conducted at the University of Konstanz, with honeybees (\textit{Apis mellifera}) from freely foraging colonies hosted on the roof. The experiment in which the alarm pheromone was obtained by pulling out stingers (see "Alarm pheromone" below) was conducted between May and August 2018. Three colonies contributed equally to this experiment ($n=252$ bees per colony, 756 bees in total). The experiment with synthetic alarm pheromone was conducted in May/June 2019. The bees were taken from 4 colonies (including one from 2018, the other colonies were lost during the winter), again in equal numbers ($n=96$ bees per colony, 384 bees in total).
To catch the bees, a black ostrich feather was waved in front of the hive entrance. The bees attacking the feather (and thus involved in colony defence) were collected in a plastic bag and cooled in ice for a few minutes, until they stopped moving. They were then placed alone in a modified syringe with ad libitum sugar water ($50\%$ vol/vol) and given at least 15 min to recover from the cold anaesthesia before testing.

\subsubsection{Alarm pheromone}

The alarm pheromone of honeybees includes over 40 compounds \citep{Collins82, Collins83}, making it difficult to synthesize. Hence in a first experiment, we pulled stingers out of cold anaesthetized bees to get the full alarm pheromone blend. This manipulation was done as fast as possible and just before the start of the trial. The stingers were placed on the dummy (the stinging target) to mimic other bees stinging it before the start of the trial. The range of alarm pheromone concentration was obtained by varying the number of stingers: 0, 1, 2, 3, 5 or 7 stingers. A clean air flow entered the arena from 3 holes on the sides, equally spaced, and the arena lid was drilled with an array of small holes to allow the air out.
The advantage of pulling stingers was to obtain the full odour blend, but the inconvenient was that the concentrations we reached were limited. To cope with this issue, we performed a second set of tests in which we only used the main component of the alarm pheromone, iso-amylacetate (IAA, Merck$\circledR$). IAA is sufficient to reproduce most of the action of the full blend \citep{Boch62, Free68}. In this case, IAA was diluted in mineral oil (Merck$\circledR$) to a final concentration (vol/vol) of: $0\%$ (control), 0,1$\%$, 1$\%$, 5$\%$, 10$\%$, 25$\%$, 50$\%$ or 100$\%$ (pure). To deliver the odour, $10 \mu l$ of solution were put on a small filter paper which was then placed inside the air flow entering the arena.
To verify that the odour concentration was linearly correlated to either the number of stingers or the dilution ratio, measurements were made inside the arena with a photoionization detector (PID). The measures were taken every 0.01 s, and the data was smoothed on a sliding 1 s (101 points) time window centered on each point. The amplitude of the odour signal was then calculated by subtracting the baseline (average of the 5 s just before the stingers were inserted or the air flow was started) from the peak value (average of the 5 s centered on the maximum value reached). For tests with IAA, a single measure was taken for each concentration. In tests with stingers, the concentrations were close to the limit of the PID sensitivity, hence we repeated the measures 3 times and averaged the results to increase reliability.

\subsubsection{Stinging assay}

The protocol for the stinging assay has been described in detail in \citep{Nouvian15}. Briefly, the bee was introduced into a cylindrical testing arena where it faced a rotating dummy coated in black leather. The stinging behaviour was first scored visually, and defined by the bee adopting the characteristic stinging posture: arched with the tip of the abdomen pressed on the dummy. This was further confirmed at the end of the trial by the presence of the stinger, embedded in the leather. 

\subsubsection{Comparison between Africanized and European bees} \label{SUBSEC AHBvsEHB}

It has been shown that Africanized bees are more sensitive to SAP \citep{Collins87}. In this study, the responses of caged honeybees to different concentrations of alarm pheromone were classified into 5 categories according to their intensity: "no response" (N), "weak response" (W), "moderate response" (M), "strong response" (S) or "very strong response" (V). To better visualize this data and to be able to compare it to our model results, we transformed this data by calculating an "Aggressive score" ($A_{s}$) for each alarm pheromone concentration and for each ecotype, which was defined as $A_{s}=N\times0+W\times1+M\times2+S\times3+V\times4$, with each letter corresponding to the percentage of reactions that fell into the corresponding category.

\subsection{Theoretical model: Projective Simulation}\label{SUBSEC projective simulation}

Projective Simulation (PS) \citep{BriegelCuevas12,Mautner15,Makmal16,Melnikov17,Melnikov18,RiedEva19,hangl2020exploratory} is a model for artificial agency that combines a notion of episodic memory with a simple reinforcement learning mechanism. It allows an agent to adapt its internal decision making processes and improve its performance in a given environment. PS has a transparent structure than can be analyzed and interpreted throughout the learning process. This feature is of particular importance in this work, since we aim at \textit{explaining} the experimentally observed individual responses to certain stimuli. 

In the context of behavioral biology, the model of PS offers the possibility of enriching the description of the entities' sensorimotor abilities to get closer to the real mechanisms, which can help gain new insight into phenomena that too simplified or abstract models cannot account for. Honeybees offer an interesting opportunity for PS since they exhibit complex behaviors at both the individual and the collective level despite their relatively small brain. In addition, Projective Simulation can be used to model collective behavior \citep{Ried19,Lopezincera20} by considering ensembles of PS agents that interact with each other. In the present work, this interaction is determined by the olfactory perception of the pheromone that bees release when stinging. The fact that each agent has an individual deliberation process allows us not only to explain the experimental results but also to study how the individual responses to alarm pheromone are combined into an appropriate defensive reaction for the colony. 

In this section, we describe the general features of Projective Simulation and we further specify how we model the scenario of colony defense in Secs.~\ref{SUBSEC Details of the model I} and~\ref{SUBSEC details of the model:predator}. 

The individual interaction of a PS agent with its surroundings starts with the agent perceiving some input information, which triggers a deliberation process that ends with the agent performing a certain action. The deliberation process is carried out by the main internal structure of the agent ---called episodic and compositional memory (ECM) \cite{BriegelCuevas12}---, which is a network consisting of nodes, termed \textit{clips}, connected by edges. Clips represent snippets (or "episodes") of the agent's experience and can encode information from basic percepts, like a color or an odor, to compositions of short sequences of sensory information. Each clip is connected to its neighboring clips by directed, weighted edges. The weights, termed $h$ values, are stored in the so-called $h$ matrix and in turn determine the transition probability from one clip to another. The deliberation process is thus modeled as a random walk through the clip network. The ECM has a flexible structure that may consist of several layers and that can change over time by, for instance, the creation of new clips and their addition to the existing network (see e.g. \citep{Melnikov17,RiedEva19}). However, for the purpose of this work, it is sufficient to consider the basic two-layered structure (see Fig.~\ref{FIG ECM interaction} (a)), where one layer of \textit{percept-clips} (or just "percepts", for brevity) encodes the perceptual information that the agent gets from its surroundings, and another layer of \textit{action-clips} (or just "actions") encodes the information about the possible actions the agent can take. 

The interaction round of an individual PS agent goes as follows: first, it perceives certain input information that activates the corresponding percept-clip in the ECM, which triggers a random walk through the network that ends when an action-clip, and subsequently its corresponding actuator, are activated, leading the agent to actually perform the action. Therefore, the final action that the agent will execute depends on the transition probabilities from clip to clip in the ECM, which are determined by the $h$ values as,
\begin{equation}
    p_{ij}=\frac{h_{ij}}{\sum_k h_{ik}}, \label{EQ probabilities}
\end{equation}
where the transition probability from clip $i$ to clip $j$, $p_{ij}$, is given by the weight $h_{ij}$ of the edge that connects them, normalized over all the possible outgoing transitions to clips $k$ connected to $i$. In this work, we consider the two-layered PS, where each percept clip is only connected to all the action clips, so $p_{ij}$ is simply the transition probability from percept $i$ to action $j$. 

A reinforcement learning mechanism can be implemented by updating the $h$ values at the end of an interaction round. If the agent's choice is good, it receives a reward $R$ that increases the $h$ value of the traversed percept-action edge, so that the agent has higher probability of performing that action the next time the same percept clip is activated. At the beginning of the learning process, we consider that the agent chooses one of the possible actions at random. Therefore, all edges are initialized with the same $h$ value, $h_{ij}^{(0)}=1$, which leads to a uniform probability distribution over the actions. In addition to an increase of the edge weights throughout the learning process, noise can be added by introducing a \textit{forgetting} parameter $\gamma$ ($0 \leq \gamma < 1$) that quantifies how much the $h$ values are damped towards their initial value. This can be interpreted as the agent forgetting part of its past experience. The specific update rule at the end of the interaction round for an edge connecting percept $i$ to action $j$ has the form,
\begin{equation}
    h_{ij}^{(t+1)} \longleftarrow h_{ij}^{(t)} - \gamma (h_{ij}^{(t)}-h_{ij}^{(0)}) + R, \label{EQ update rule w/o glow}
\end{equation}
where $h_{ij}^{(t)}$ denotes the current $h$ value, $h_{ij}^{(0)}$ the initialized $h$ value at the beginning of the learning process and $R \geq 0$ the reward. If the transition from percept $i$ to action $j$ is rewarded, then $R$ has a value $R>0$, whereas if it is not rewarded, $R=0$ and the edge weight is only damped. Note that this update rule increases the $h$ value at the end of each interaction round, depending on whether a reward is given for that round. If one considers a scenario where the agent interacts for several rounds and only gets a reward at the end of the last one, then only the percept-action edge that is traversed in that round is enhanced. In order to reinforce all the percept-action pairs that led to a reward, a mechanism called \textit{glow} is introduced as part of the model. The idea is to keep track of which edges are traversed during the interaction rounds before the reward is given. To do so, once an edge is traversed, a certain level of excitation or "glow" is associated to it with the effect that, when the \emph{next} reward is given, all the "glowing" edges are enhanced in proportion to their glow level. The glow for each edge $i \rightarrow j$ is stored in the element $g_{ij}$ of the so-called glow matrix $g$, and the update rule of eq.~\eqref{EQ update rule w/o glow} takes the form,
\begin{equation}
    h_{ij}^{(t+1)} \longleftarrow h_{ij}^{(t)} - \gamma (h_{ij}^{(t)}-h_{ij}^{(0)}) + g_{ij} R. \label{EQ update rule w glow}
\end{equation}
Therefore, if that edge is glowing ($g_{ij}=1$) at the end of the interaction round where the reward is given, it will be enhanced. For the purpose of this work, it is sufficient for the reader to consider that edges get a glow value $g_{ij}=1$ if they are traversed and $g_{ij}=0$ if they are not. For more details on how to assign and update glow values in different scenarios, we refer the reader to \citep{Mautner15,Boyajian19}.

So far, we have described the main processes that a single PS agent carries out when interacting with its surroundings and learning via reinforcement. In this work, we model a scenario with an ensemble of PS agents, each of which has its own deliberation process and makes decisions independently from the rest of the ensemble. The precise details of how the agents interact with each other and how the collective performance is evaluated are given in Sec.~\ref{SUBSEC Details of the model I}. There are a few remarks still to be made regarding the learning process of such an ensemble and its interpretation from a biological point of view. In this work, we do not assume that the individual biological entities have the capacity to learn, but we consider the learning process from an evolutionary perspective. Hence, the improvement of the collective performance of the ensemble of agents throughout the learning process can be interpreted as the adaptation of a given species to certain pressures throughout its entire evolutionary history. In the context of reinforcement learning, the selection pressure is encoded in the reward function, in such a way that we can test hypotheses about which environmental factors may have influenced the resultant behavior we currently observe in the real organisms. In this view, the forgetting parameter would capture one aspect of genetic drift.

\subsection{Details of the model I: The bee and the learning process}\label{SUBSEC Details of the model I}
We consider a population of $G$ bees, where each bee is modeled as a PS agent that perceives, decides, acts and learns according to the model explained in the previous section. The population is confronted with the pressure of a predator that attacks the colony and kills agents until it is scared away, i.e. until it is stung a certain number of times. In this section, we describe how we model the colony and the learning process. We give further details on the model of the predator in Sec.~\ref{SUBSEC details of the model:predator}.

Honeybees release an alarm pheromone when their stinger is exposed, which allows them to alert and recruit nearby bees into a collective defensive response. Since we are interested in explaining the experimentally observed response of bees to the sting alarm pheromone, we consider that the agents decide whether to sting or not based on the pheromone concentration they perceive. The alarm pheromone concentration is discretized in our model and it increases by one unit every time an agent decides to sting. Hence, the minimum concentration is 0 units, when no agent stings, and the maximum is $G$ units, if all agents sting. Note that each agent can only sting once, like a real bee (when a bee stings an animal with elastic skin such as a mammal or bird, its stinger remains hooked into the predator thanks to its barbs, and tears loose from the bee's abdomen).

\begin{figure}[htb!]
\includegraphics[width=3.4in]{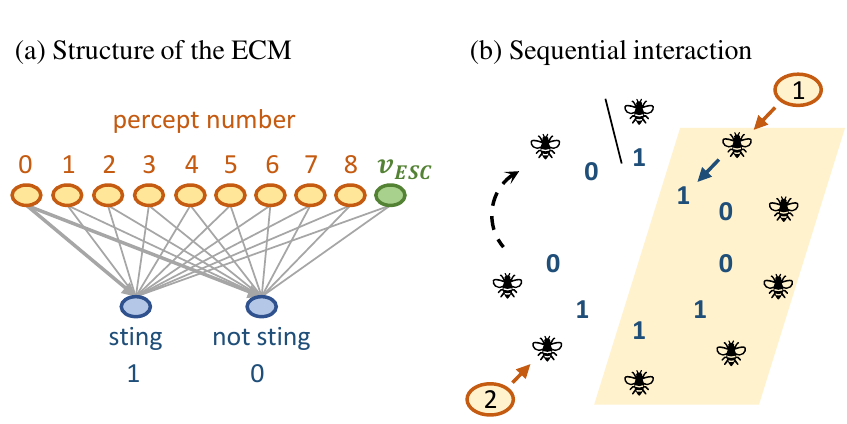}
\caption{(a) Structure of each agent's ECM. There are 10 percepts (logarithmic binning of pheromone units and one visual stimulus "$v_{\text{ESC}}$") and 2 actions (sting or not sting). See text for details on the perceptual apparatus of our agents. (b) Sequential interaction (trial) of one population (it starts at the line and continues clockwise in the picture), which comprises $G$ time steps. In this example, we consider that $G=100$, the correspondence between percept number and pheromone units is given in Table~\ref{TAB:logbinning} and the chosen action is depicted in blue (1 for stinging and 0 for not stinging). Thus, the agents within the shaded area perceive the same input information (percept 1), namely that the pheromone concentration is in the range between 1 and 3 units. The seventh agent perceives a change in concentration (percept 2).}\label{FIG ECM interaction}
\end{figure}

In order to define the structure of the ECM and the percepts of the PS model, we group the range of $G$ pheromone units into bins of logarithmic size, where each bin corresponds to one percept (see Table~\ref{TAB:logbinning} and Fig.~\ref{FIG ECM interaction} (a)). There are two reasons for this choice. First, the logarithmic sensing implies that the agents are able to resolve with high precision low values of pheromone concentration but that precision decreases as the concentration increases. On the one hand, with a range of pheromone units extending to $G\sim 10^2$, it is plausible to assume that the agent can distinguish between 0 and 1 pheromone units but a change by 1 pheromone unit becomes harder to sense when the concentration is of the order of e.g. 30 units. Furthermore, several species of animals present a logarithmic relationship between the stimuli and their perception, as it is the case for instance with human perception, which is quantified by the Weber-Fechner law. The second reason for this choice is related to the structure of the PS agents' interaction. We consider the collective interaction to be sequential, i.e. the agents perceive, decide and act one after the other until all of the $G$ agents have made their decision and acted. This \emph{a priori} artificial condition becomes more realistic with the logarithmic binning since several agents perceive the same percept and hence act based on the same input information, which is effectively as if they would react in parallel. An illustrative example of the sequential interaction, named trial, is given in Fig.~\ref{FIG ECM interaction} (b). Each trial is always a defensive event that is triggered by an unspecified disturbance outside the nest. Each of the agents decides only once whether to sting or not, hence the trial comprises $G$ time steps, where a "time step" is defined to be the decision time of one agent.

\begin{table}[htb]
\begin{tabular}{|c|c|}
\hline
Percept number & Ph. units \\ \hline
0 & 0\\ \hline
1 & 1..3\\ \hline
2  & 4..6 \\ \hline
3 & 7..9 \\ \hline
4 & 10..15 \\ \hline
5 & 16..25 \\ \hline
6 & 26..39 \\ \hline
7 & 40..63 \\ \hline
8 & 64..99 \\ \hline
\end{tabular}
\caption{Correspondence between the percept number and its range of pheromone units for a group of size $G=100$ bees, providing that the binning is logarithmic. 10 bins are initially considered and bins with less than 3 units are merged together to make the learning process more stable. Percept 0 is reserved for the initial situation where there is no pheromone yet.}
\label{TAB:logbinning}
\end{table}

Since we want to model a situation as close as possible to the experimental one, we assume that our agents also have a visual perception of the predator during the whole process (the bees see a rotating dummy, as explained in Sec.~\ref{SUBSEC exper. methods}). As in the experimental setup, we consider that this visual perception remains unchanged while the agent perceives the increase in the pheromone concentration (percepts 0 to 8, see Table~\ref{TAB:logbinning}). However, in our model we add an additional percept (labeled as "$v_{\text{ESC}}$" in Fig.~\ref{FIG ECM interaction} (a)) that is activated when the visual perception of the predator changes and the agents \textit{see} that the predator is already escaping and is no longer a threat to the colony. In this case, we assume that the visual input overrides the olfactory one so that the agents decide based on the visual information only. This is consistent with reports that show that while the alarm pheromone recruits bees to the location of the disturbance, a moving visual stimulus is then necessary to release the stinging behavior \citep{Wager00}. In order to study the interplay between the visual and the olfactory information and their influence on the self-limitation of the stinging response, we consider the percept "$v_{\text{ESC}}$" to not be activated immediately after the predator stops attacking, but after some time delay $\Delta t_v$. This time delay is more realistic than an immediate cut-off: it corresponds to the time needed for the bees to perceive that the predator is retreating. In addition, the predator may need time to move outside of the perimeter defended by the resident bees.

In order to avoid unnecessary losses, we could expect the agents to learn to stop stinging already during the interval $\Delta t_v$, based on the pheromone concentration (i.e. on the number of bees that already stung, which may be sufficient to deter any predator). We can thus analyze the two extreme cases with $\Delta t_v=0$ and $\Delta t_v=\infty$, which correspond respectively to when bees are able to immediately distinguish that the predator is leaving ($\Delta t_v=0$) or to when they completely ignore this visual cue ($\Delta t_v=\infty$). In the former case, we expect no self-limiting mechanism to develop, since the end of the attack is always immediately signaled by the visual stimulus. In the latter scenario however, unusually high alarm pheromone concentrations may start playing the role of this "stop signal". Note that in the experimental setup the dummy is always rotating, hence the bees' reactions are modulated by the pheromone level exclusively.

With respect to the learning process, the performance of the population is evaluated at the end of one sequential interaction (or \textit{trial}) and the individuals are rewarded or not depending on the final state of the colony. Thus, the agents' ECMs do not change during the trial. At the end of it, a reward that is proportional to the number of surviving agents is given and the ECMs of the agents are updated accordingly. Importantly, the same $h$ matrix is used for every agent of the population. The reason for this choice is that genetic mixing during reproduction means that, effectively, the average individual responses to the alarm pheromone are transmitted to the next generation \citep{Avalos20}. One could easily imagine that a scenario in which certain agents always sting while others never do may lead to a viable collective strategy, but such a population structure could not be stably maintained across generations.
As explained in Sec.~\ref{SUBSEC projective simulation}, the update rule of eq.~\eqref{EQ update rule w glow}, now given in matrix form, reads,
\begin{equation}
h^{(t+1)}=h^{(t)}-\gamma \, (h^{(t)}-h^{(0)}) + g R.
\end{equation}
where $h^{(t)}$ is the $h$ matrix at the current trial and $h^{(0)}$ denotes the initial $h$ matrix (a $10\times 2$ matrix with ones in all its entries). Note that agents start with a probability of stinging $p_s=0.5$ for all the percepts. The already learned responses are damped by a factor $\gamma$ at the end of each trial. The influence of this parameter on the learning process is further studied in Sec.~\ref{SUBSEC learning process}. In this work, we adapt the notion of a glow matrix $g$ presented in Sec.~\ref{SUBSEC projective simulation} to take into account the choices of all agents and distribute the reward depending not on the individual performance but on the collective one. We remark that the learning process is, in our case, interpreted as the evolutionary history of honeybees. Therefore, even though there exist fluctuations at the individual level, we are interested in the average effect on the population. From this perspective, we consider a glow matrix $g$ that stores how many agents chose an action given a certain percept. In the example of Fig.~\ref{FIG ECM interaction} (b), the second row of $g$ ---the one corresponding to percept 1--- in that trial is (2,3), which indicates that 2 agents decided not to sting and 3 decided to sting. If the population is rewarded at the end of the trial, the individual responses that lead to a good collective performance are enhanced. For instance, if the optimal defense is that all the bees sting from the beginning, the individual probability of stinging for low pheromone concentrations will converge to a high value. 


As to the reward, it is determined by the percentage of bees that remain alive at the end of the trial,
\begin{equation}
R=\frac{a}{G}, \label{EQ reward function}
\end{equation}
where $a$ denotes the number of live bees. This number is evaluated at the end of each trial, by subtracting the number of dead bees from the total number of agents $G$. Bees die after stinging ($s$) or because they are killed by the predator ($q$),
\begin{equation}
a=G-s-q.
\end{equation}
Note that the number of agents does not change during the trial (all G agents get to decide and act, bees killed by the predator are only counted at the end of the trial). This choice of reward system reflects the fact that honeybee colonies with a larger workforce are more successful ecologically: they have a better chance of surviving winter \citep{Doke15}, and most importantly they are more likely to be able to invest in reproduction in spring \citep{Smith14}.

In the simulations reported here, the entire learning process consists of $80,000$ trials, which is sufficient for the population to converge to a stable behavior.

We remark that the learning processes that the populations of PS agents undergo are interpreted as processes of adaptation to given evolutionary pressures. In this work, we focus on the defense behavior against predators, so, by changing the parameters of our predator model, we are effectively testing how different pressures affect the final behavior. This allows us to analyze possible causal explanations 
for the responses observed in present-day real bees.

\subsection{Details of the model II: The predator}\label{SUBSEC details of the model:predator}
\begin{figure}[htb!]
\includegraphics[width=3.4in]{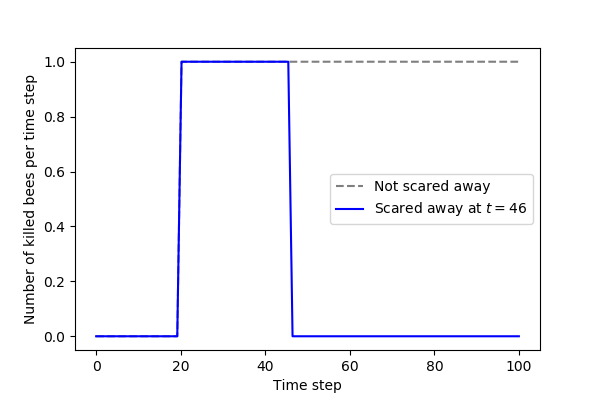}
\caption{Predator's behavior during one trial with a colony of 100 bees. The number of killed bees is depicted as a function of time. In this particular example, the predator starts killing bees at $t_{att}=20$ and kills $k=1$ bee per time step. The dashed line represents its behavior if it is not scared away and the continuous line represents the case where it is scared away and stops killing.}\label{FIG predators action}
\end{figure}

The predator has an active role in our model, since it attacks the hive and kills bees at a given killing rate of $k$ bees per time step. Therefore, the colony needs to build up the defense as fast as possible to reduce the number of bees killed by the predator. Of particular importance is the time needed for the bees to detect the presence of the predator, which we parametrize as the time step at which the predator starts its attack $t_{att}$ (see Fig.~\ref{FIG predators action}). A low value for this parameter simulates cases in which the predator is only detected close to the colony, and hence starts killing bees quickly. At the opposite, a high value for $t_{att}$ represents an early detection by the bees, when they have more time to fly out and build up the defensive response before the predator reaches the nest itself.

The predator stops killing bees when the number of total stings reaches a threshold $s_{th}$. By changing this parameter, we model the type of predator that the colony may encounter. As an example, one may consider that small predators such as mice can be killed or scared away with fewer stings than bigger or thick-skinned animals like bears and honey badgers. Thus, different $s_{th}$ can be interpreted as differences in the predator's resistance to bee stings.

In the wild, bees regularly encounter a wide variety of predators, and they need to be able to cope with all of them. We model this situation by introducing a range of $s_{th}$. Instead of being faced with only one type of predator (same $s_{th}$ in all trials), the colony is attacked by a predator with a different $s_{th}$ for different trials, which is chosen from a uniform distribution over a certain range. Therefore, the parameter $s_{th}$ gives us the flexibility to model different environmental conditions. For instance, we can model colonies of bees that are usually attacked by small/less resistant predators and observe the defensive strategy that they adopt. We can then study how they respond when suddenly faced with bigger/more resilient predators, thus mimicking their introduction into a novel environment. 

Since the agents can only develop one strategy (i.e. a set of probabilities of stinging for the various percepts) per learning process, they have to optimize it to accommodate the whole range of $s_{th}$. Note that the activation of the visual percept "$v_{\text{ESC}}$" ---which happens when the visual information changes and the predator is seen leaving--- allows them to stop the defense behavior at different points of the trial depending on the specific $s_{th}$ of each predator. In particular, percept "$v_{\text{ESC}}$" is activated after $\Delta t_v$ time steps from the point where the number of stings reaches the corresponding $s_{th}$.
 
The effect of all the aforementioned predator's parameters on the collective behavior that agents develop is studied in Sec.~\ref{SUBSEC predators parameters}.

\section{Results}
\subsection{Experimental results} \label{SUBSEC Experimental results}
\begin{figure}[htb!]
\includegraphics[width=3.4in]{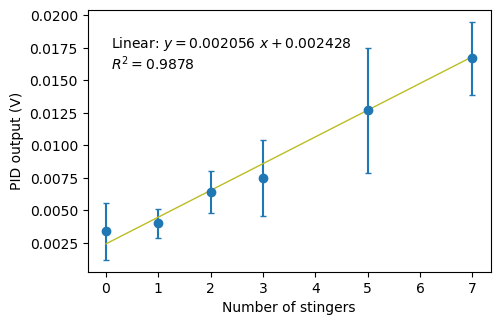}
\caption{PID measurements of SAP concentrations in the arena depending on the number of pulled stingers. Mean of 3 measurements $\pm$ standard deviation.}\label{FIG exp curve 1}
\end{figure}

\begin{figure}[htb!]
\includegraphics[width=3.4in]{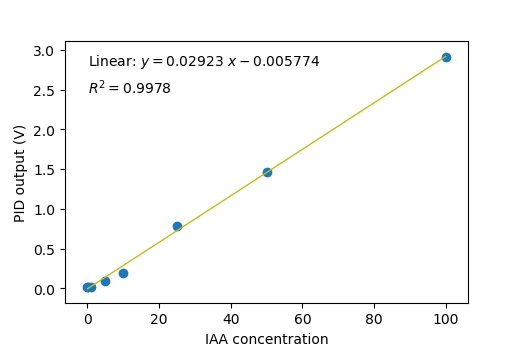}
\caption{PID measurements of IAA concentrations inside the arena.}\label{FIG exp curve 2}
\end{figure}

To better understand if and how bees use the alarm pheromone concentration as a source of information during a defensive event, we first sought to establish a dose-response curve to the SAP. This requires to precisely control the pheromone concentration inside the testing arena. We used two methods to create a range of alarm pheromone concentrations: 1) pulling out a defined number of stingers from live, cold-anaesthetized bees, or 2) diluting synthetic IAA. To verify that the final concentrations scaled linearly with either the number of stingers or the dilution factor, we realized PID measurements inside the arenas. As shown on Figures~\ref{FIG exp curve 1} and~\ref{FIG exp curve 2}, both methods reliably created linear series of concentrations (IAA: Pearson's $r=0.9989$, $p<0.001$; stingers: Pearson's $r=0.9922$, $p<0.001$). The absolute concentrations reached by using stingers appear to be much lower than the ones obtained with synthetic IAA, but one should keep in mind that the delivery method was also very different (stingers placed on the dummy vs. IAA carried in by the air flow) and that only a subset of the SAP compounds can be detected by the PID, making a direct comparison difficult.

\begin{figure}[htb!]
\includegraphics[width=3.4in]{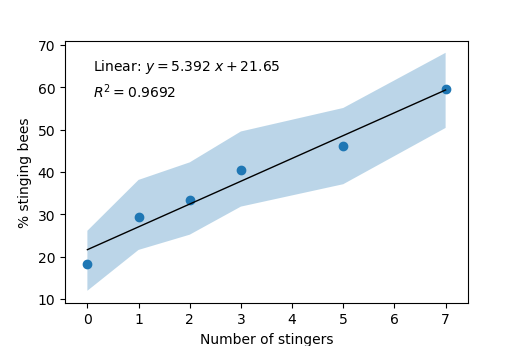}
\caption{Stinging frequency of a bee as a function of the number of pulled stingers on the dummy. The shaded area corresponds to the 95\% confidence interval, estimated from a binomial distribution with our sample size (126 bees for each point).}\label{FIG exp curve 3}
\end{figure}

\begin{figure}[htb!]
\includegraphics[width=3.4in]{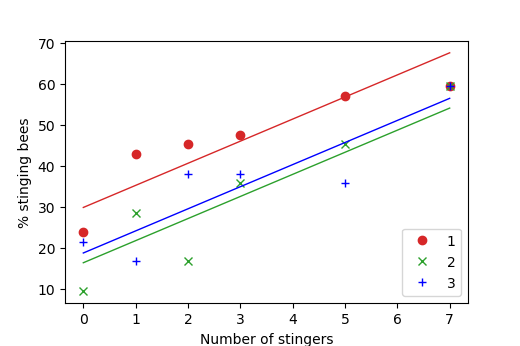}
\caption{Stinging frequency of a bee as a function of the number of pulled stingers on the dummy and its colony of origin.}\label{FIG exp curve 4}
\end{figure}

We first evaluated the stinging response of single bees faced with a rotating dummy on which a certain number of freshly pulled stingers were placed, to mimic previous attacks by other bees. The stinging likelihood of an individual bee increased linearly with the number of stingers (see Fig.~\ref{FIG exp curve 3}), from about 20\% of the bees reacting to the dummy alone to 60\% of them stinging when 7 stingers –--and hence 7 "units" of SAP--– were added ($n= 126$ bees per data point so 756 bees in total; Pearson's $R=0.9845$, $p<0.001$). Three colonies contributed equally to this dataset, which allowed us to check for variations in this response pattern (see Fig.~\ref{FIG exp curve 4}). We found no significant difference on the regression slope between colonies (ANOCOVA, interaction term: $F(2,754)=0.9$, $p=0.432$), indicating that the effect of the SAP was similar on all bees. However, bees from colony 1 were overall more likely to sting than bees from colony 2 (ANOCOVA followed by Tukey's HSD on intercepts, $p=0.024$), while bees from colony 3 showed intermediate aggressiveness (ANOCOVA followed by Tukey's HSD on intercepts, 1 vs. 3: $p=0.150$, 2 vs. 3: $p=0.550$). Such behavioural variability is not surprising, as it is known that the aggressiveness of honeybees is strongly influenced by genetic factors \citep{Hunt07}.

\begin{figure}[htb!]
\includegraphics[width=3.4in]{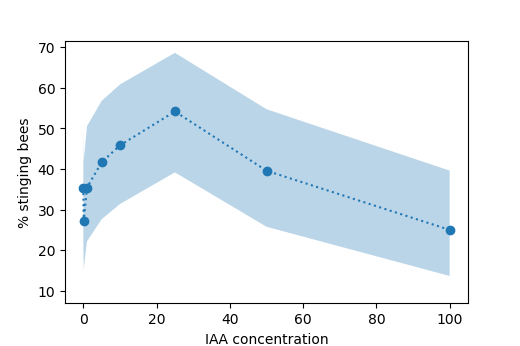}
\caption{Stinging frequency of a bee as a function of the IAA concentration inside the arena. The shaded area corresponds to the 95\% confidence interval, estimated from a binomial distribution with our sample size (48 bees for each point).}\label{FIG exp IAA response}
\end{figure}

The advantage of getting the SAP from stingers is that we could work with the full pheromonal blend, which is otherwise difficult to obtain. Its main inconvenience, however, is that only limited concentrations can be reached. To get at these higher concentrations, we repeated the experiment with different dilutions of IAA, the main active component of the SAP. The results are shown in Fig.~\ref{FIG exp IAA response}. Consistent with our previous results, we observed a linear increase in stinging responsiveness between 0 and 25\% IAA (Pearson's $r=0.9127$, $p=0.011$). The two higher concentrations that we sampled (50\% and 100\%) revealed a decay in aggressiveness. While these 3 points were not sufficient to test for a correlation, the decrease in stinging frequency between 25\% and 100\% was significant ($X2(1,96)=7.3612$, $p=0.007$).

\subsection{Theoretical results I: Parameter tuning} \label{SUBSEC learning process}
In this section, we study the learning process and how learning parameters ---also called hyperparameters--- such as forgetting ($\gamma$) influence the global performance. To this end, we set up a task that admits a clear optimal collective strategy that can be easily predicted beforehand, allowing us to calibrate the learning process with a suitable $\gamma$. We consider just one predator with $s_{th}=26$ and $t_{att}=0$, for which the optimal strategy is known: the first $26$ bees should sting with $p_s=1$ and the remaining ones should not sting ($p_s =0$). In this case, if the predator kills one bee per time step, the number of live bees at the end of the trial is maximized ($a^*=100-2\cdot 26=48$). Note that we have chosen $s_{th}=26$ so as to allow the bees to distinguish when they have already stung exactly $26$ times, since this is the start of a new binning interval that corresponds to a different percept (percept $6$). 

\begin{figure}[htb!]
\includegraphics[width=3.4in]{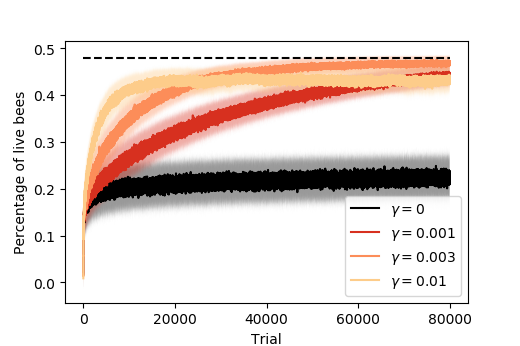}
\caption{Evolution of the percentage of live bees through the learning process for different values of $\gamma$. Dashed line indicates the optimal value of 0.48. Average is taken over 50 independently trained populations of 100 agents. Shaded areas depict one standard deviation. Parameters: $s_{th}=26, k=1, t_{att}=0, \Delta t_v=10$.}\label{FIG learning gamma}
\end{figure}

We study how the learning process for this task is influenced by the choice of the forgetting parameter $\gamma$. Fig.~\ref{FIG learning gamma} shows the evolution of the percentage of live bees through the learning process for different values of $\gamma$. If the forgetting is too low ($\gamma=0$) or too high ($\gamma=0.01$), the agents' behavior converges quickly but gets stuck in local minima, and the optimal performance $a^*$ is not reached. The intermediate value $\gamma=0.003$ enables the agents to slowly improve their performance until they reach a percentage of live bees $a=46.9\pm 1.5$ (see also Table~\ref{TAB:performance forgetting}).

\begin{figure}[htb!]
\includegraphics[width=3.4in]{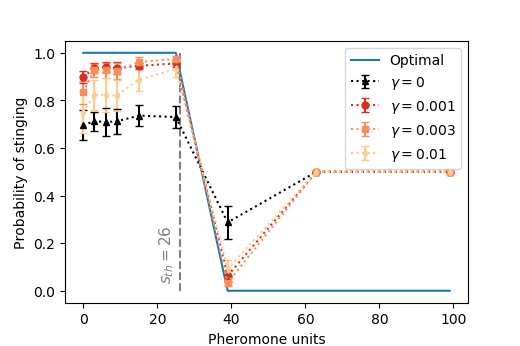}
\caption{Learned probability of stinging as a function of the pheromone units, for different $\gamma$. Average and standard deviation are obtained by taking data at the end of the learning process of 50 independently trained populations of 100 agents. Parameters: $s_{th}=26, k=1, t_{att}=0, \Delta t_v=10$.}\label{FIG policy forgetting}
\end{figure}

Fig.~\ref{FIG policy forgetting} shows the learned probability of stinging as a function of the pheromone units for cases with $\gamma=0, 0.001, 0.003, 0.01$. The behavior that is closest to the optimal one is obtained for $\gamma$ with order of magnitude $10^{-3}$. For $\gamma=0.01$, the probability of stinging at low pheromone concentrations is lower than for $\gamma=0.001, 0.003$, which leads to a slower collective defensive response. This allows the predator to kill more bees before it is scared away (see Table~\ref{TAB:performance forgetting}), so the final performance of populations that learned with $\gamma=0.01$ is lower than those with $\gamma=0.001, 0.003$ (see Fig.~\ref{FIG learning gamma}). Note that the last two percepts, corresponding to pheromone units up to 63 and 99, respectively, retain the initial values $p_s=0.5$. This is due to the fact that, once agents learn to stop stinging at percept 6 (26-39 pheromone units), they do not encounter those two percepts any more and their values $p_s$ do not change (in case these higher-concentration percepts were only encountered earlier in the learning process, their value $p_s$ decays back to $0.5$ due to the effect of forgetting). This implies that these two probabilities do not affect the collective performance at all. 

When analyzing the probability of stinging for percept "$v_{\text{ESC}}$", we observe that, as expected, agents learn to stop stinging ($p_s \rightarrow 0$). The obtained values are $p_s=0.25 \pm 0.03$, $p_s=0.042 \pm 0.007$, $p_s=0.006 \pm 0.003$ and $p_s=0.015 \pm 0.007$ for processes with $\gamma=0, 0.001, 0.003, 0.01$, respectively. Note that populations trained with $\gamma=0.003$ achieve the lowest value, which further supports our finding that this order of magnitude for the forgetting parameter is suitable.

\begin{table}[htb]
\begin{tabular}{|l|c|c|c|c|}
\hline
& $\gamma=0$ & $\gamma=0.001$ & $\gamma=0.003$ & $\gamma=0.01$ \\ \hline
Stinging & $42.5\pm 4.0$ & $29.3\pm 1.8$ & $26.7\pm 0.9$ &  $27.7\pm 1.4$\\ \hline
Killed  & $35.1\pm 3.9$ & $26.6\pm 1.3$ & $26.4\pm 1.3$ & $28.9\pm 2.4$ \\ \hline
Alive ($a$) & $22.4\pm 4.7$ & $44.2\pm 2.2$ & $46.8\pm 1.5$ & $43.3\pm 2.7$  \\ \hline
\end{tabular}
\caption{Performance achieved at the end of the learning process for different values of $\gamma$. Averages are taken over the last 500 trials and over 50 independently trained populations of 100 agents.}
\label{TAB:performance forgetting}
\end{table}

\begin{figure}[htb!]
\includegraphics[width=3.4in]{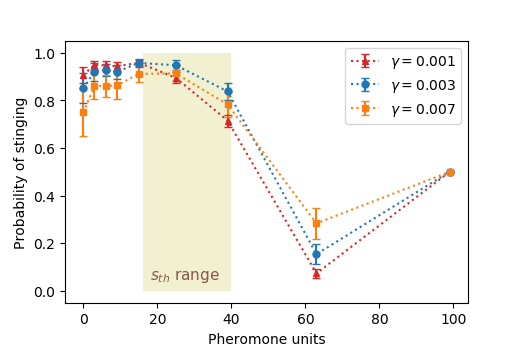}
\caption{Probability of stinging as a function of the pheromone units at the end of a learning process with a range of predators with $s_{th}\in (16,40)$ (depicted as a shaded area), for different values of $\gamma$. Average and standard deviation are obtained by taking data at the end of the learning process of 50 independently trained populations of 100 agents. Parameters: $k=1, t_{att}=0, \Delta t_v=10$.}\label{FIG policy calibration sth range}
\end{figure}

The forgetting parameter is also crucial in tasks with changing objectives, for instance when different predators attack the colony (i.e. when $s_{th}$ can vary throughout the trial). We analyze the effect of $\gamma$ on such learning processes in the following. In particular, we consider a task where $s_{th}$ is chosen from a uniform distribution over the range $s_{th}\in (16,40)$. We remark that the agents should develop a single strategy that is suitable for all values of $s_{th}$ within the same learning process, that is, there is no independent learning process for different values of $s_{th}$. Fig.~\ref{FIG policy calibration sth range} shows the learned probabilities of stinging for processes with $\gamma=0.001, 0.003, 0.007$, where we observe that $p_s$ has a high value for pheromone concentrations below 26 pheromone units. At higher concentrations, the probability of stinging decreases gradually, reflecting the fact that only few predators necessitate that many stings before escaping. In contrast, if only one value of $s_{th}$ is considered in the learning process, the decrease of $p_s$ is more abrupt (see Fig.~\ref{FIG policy forgetting}). In addition, agents learn to stop stinging when the percept "$v_{\text{ESC}}$" is activated in all three learning processes, but the lowest value of $p_s$ is achieved for $\gamma=0.003$ (see Table~\ref{TABLE data ps for "v" (I).} of Appendix~\ref{APP prob sting for v}).

\begin{figure}[htb!]
\includegraphics[width=3in]{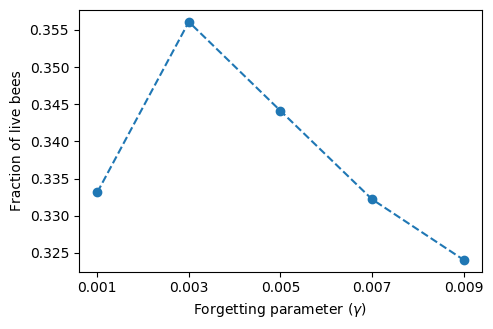}
\caption{Fraction of live bees as a function of the forgetting parameter $\gamma$ for a task with $s_{th}\in (16,40)$. Each data point corresponds to the average performance of 50 independently trained populations of 100 agents in the last 500 trials of the learning process. Parameters: $k=1, t_{att}=0, \Delta t_v=10$. }\label{FIG changing task}
\end{figure}

Analogously to these three learning processes with $\gamma=0.001, 0.003, 0.007$, we have studied the performance of colonies trained with values of $\gamma$ in the range $(0.001,0.1)$ in order to select the forgetting parameter that leads to a behavior as close as possible to the optimal one. In the context of machine learning, this calibration process is referred to as hyperparameter optimization (see \citep{Melnikov18} for PS). The results are presented in Fig.~\ref{FIG changing task}, where we see that the value $\gamma=0.003$ achieves the best performance. Therefore, we choose this value to further extend our analysis and study the influence of the predator's parameters in the following section. 

\begin{figure}[htb!]
\includegraphics[width=3.4in]{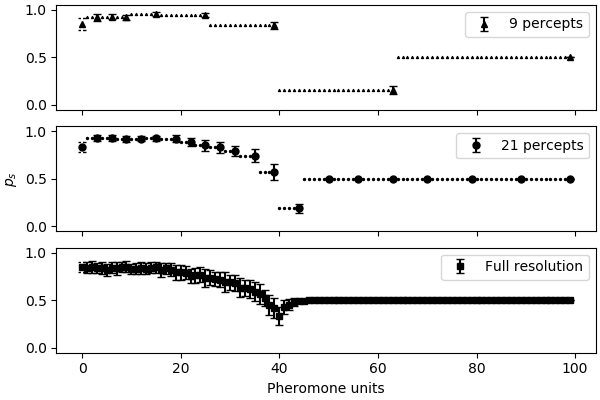}
\caption{Learned probability of stinging ($p_s$) at the end of learning processes where agents have different sensing abilities. The first two panels show processes with 9 and 21 percepts (logarithmic bins), and the last panel shows a process where agents have full resolution of the pheromone units, i.e. they are able to distinguish an increase in one unit. Average and standard deviation are obtained by taking data over 50 independently trained populations of 100 agents. Large markers are at the end of each bin. Parameters: $\gamma=0.003, s_{th}\in (16,40), k=1, t_{att}=0, \Delta t_v=10$.} \label{FIG policy resolution}
\end{figure}

Finally, we analyze the influence of the logarithmic binning and the number of percepts in the learning process, in order to verify that different choices in parametrizing the bees' perception lead to the same predicted behavior. To do so, we consider a learning process with a range of predators $s_{th}\in (16,40)$ and $\gamma=0.003$ and we vary the resolution with which the agents can perceive the number of pheromone units. Fig.~\ref{FIG policy resolution} shows the learned probabilities of stinging for processes with 9 and 21 percepts with logarithmic binning and 100 percepts with linear binning, named "full resolution", since agents are able to resolve increases by a single pheromone unit. Note that, for the logarithmic binning, we consider that the minimum size of each bin is three pheromone units. Thus, 40 percepts reduce to 21 when the first bins that have sizes 0.5, 1 etc. are combined to form one bin of size 3. We observe that the behavior that the agents develop does not qualitatively change depending on our choices in modeling their perceptual mechanisms. In all three processes, agents learn to sting with high probability when the concentration of pheromone is low and they learn to stop stinging at 40 units, which is the $s_{th}$ of the largest predator. The probability of stinging for percept "$v_{\text{ESC}}$" tends to zero in all cases (see Table~\ref{TABLE data ps for "v" (I).} of App.~\ref{APP prob sting for v}). The only difference between these 3 ways of modeling the sensory perception of the agents was the resolution achieved: more percepts (i.e. smaller bins) allowed more gradual transitions in $p_s$ values, but the results remained qualitatively similar. Therefore, in the rest of this work, we consider 9 percepts, since this makes the learning process more stable by presenting agents fewer percepts for which to learn a response and is less computationally demanding.

\subsection{Theoretical results II: Effect of the different environmental variables on honeybee responses to the SAP} \label{SUBSEC predators parameters}

In this section, we fix the learning parameter to $\gamma=0.003$ and study the effect of all the model parameters: the time at which attack starts $t_{att}$, the killing rate $k$, the range of $s_{th}$ and the time delay $\Delta t_v$ from the moment where $s_{th}$ is reached until the visual percept "$v_{\text{ESC}}$" is activated. In addition, we analyze the influence of the predation rate and the size of the defended area on the defensive behavior of the colony. All the following learning processes have been run with 50 independent ensembles of 100 agents, for 80000 trials. For each figure, mean and standard deviation are obtained by taking data at the end of the learning processes for the 50 ensembles, when their performance has converged. For clarity, percepts for which the probability of stinging remains at the initial values ($p_s=0.5$) are not shown in the figures. 

\begin{figure}[htb!]
\includegraphics[width=3.4in]{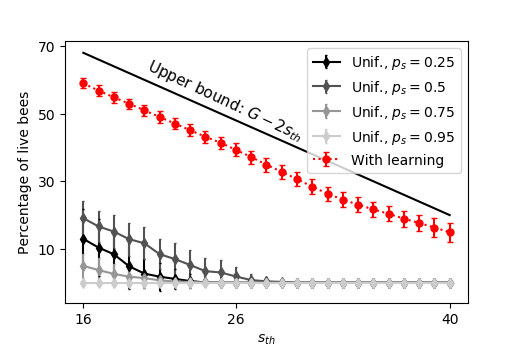}
\caption{Comparison of the performance of colonies that have gone through a learning process with colonies that follow a fixed strategy: bees stinging with probability $p_s$ independently of the pheromone concentration. Solid lines (diamonds) show the percentage of live bees for strategies with values of $p_s=0.25,0.5,0.75,0.95$. Average and one standard deviation are obtained from data of 5000 trials with no learning. Colonies that go through the learning process and develop different responses depending on the pheromone concentration keep more live bees for all the encountered predators. The upper bound indicates the optimal performance for each value $s_{th}$, which considers that exactly $s_{th}$ bees sting in the first $s_{th}$ time steps (i.e. $s_{th}$ bees are killed by the predator). Average and one standard deviation are obtained from data of the last 500 trials of the learning process. Parameters: $\gamma=0.003, s_{th}\in (16,40), k=1, t_{att}=0, \Delta t_v=10$.}\label{FIG live bees baseline}
\end{figure}

Fig.~\ref{FIG live bees baseline} shows that colonies that go through a learning process perform much better than those whose bees sting with a fixed probability for all percepts, independently of the pheromone concentration or the visual stimulus they perceive. This analysis already shows that an individual stinging response that varies depending on the pheromone concentration is indeed advantageous for the colony. The concrete responses that the bees develop throughout the learning process are studied in the following, where we also analyze how the different environmental variables shape the individual response-curve to the pheromone.

\subsubsection{Interval between predator detection and the start of its attack}
\begin{figure}[htb!]
\includegraphics[width=3.4in]{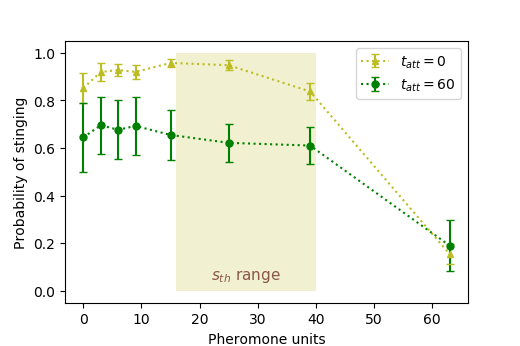}
\caption{Learned probability of stinging as a function of the pheromone units for different values of $t_{att}$. Parameters: $\gamma=0.003, s_{th}\in (16,40), k=1, \Delta t_v=10$.}\label{FIG policy comparison t}
\end{figure}

\begin{figure}[htb!]
\includegraphics[width=3.4in]{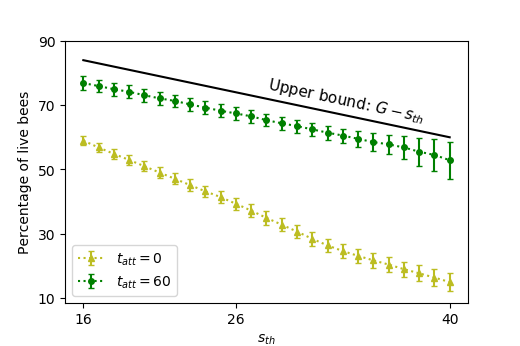}
\caption{Comparison of the percentage of live bees for different values of $t_{att}$. Solid line indicates an upper bound for the performance of the colony, which considers that exactly $s_{th}$ bees sting and no bees are killed by the predator. Average and one standard deviation are obtained from data of the last 500 trials of the learning process. Parameters: $\gamma=0.003, s_{th}\in (16,40), k=1, \Delta t_v=10$.}\label{FIG live bees comparison t}
\end{figure}

A small proportion of honeybees, termed guards, sit at the nest entrance and monitor its surroundings \citep{Nouvian16}. They may detect predators early enough to start the defensive response before the intruder reaches the colony. We wondered how the presence of guards may benefit the colony, and in particular how it may affect the later response of bees recruited via the alarm  pheromone. To address this question, we varied the time $t_{att}$ at which a predator starts killing bees after it was detected: high values for $t_{att}$ thus represent colonies that invest heavily in guards or monitor large areas, whereas low values can be taken to represent colonies that only get alerted once the predator is already close by. As shown in Fig.~\ref{FIG policy comparison t}, we find that the probabilities of stinging are lower when the bees detect the predator early ($t_{att}=60$) than when they do not have guards ($t_{att}=0$). Nonetheless, populations with guards actually fare better than their counterparts, as they always manage to deter the predator before it starts killing bees (Fig.~\ref{FIG live bees comparison t}). Note that their performance does not reach the upper bound $G-s_{th}$ because they do not stop stinging immediately after $s_{th}$ is reached, but after a time delay $\Delta t_v=10$ (see Sec.~\ref{SUBSEC defense termination}). The analysis of Fig.~\ref{FIG live bees comparison t} confirms that guards are beneficial for the colony, as we expect. Moreover, the overall lower excitability of colonies with guards likely also gives them an advantage by preventing losses in case of false alarms (see Sec.~\ref{SUBSEC Predation rate} below). In the experimental data we always controlled the alarm pheromone concentration, so the initiating role of guards was skipped. For this reason, we set $t_{att}=0$ in all the following sections.

\subsubsection{Killing rate}
\begin{figure}[htb!]
\includegraphics[width=3.4in]{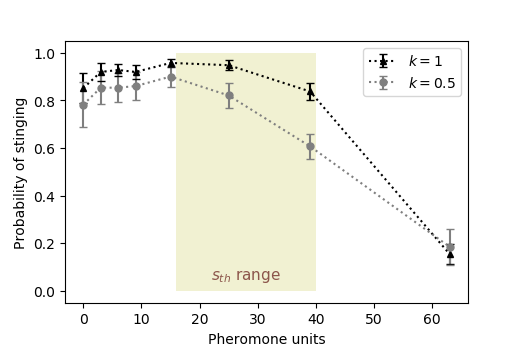}
\caption{Learned probability of stinging as a function of the pheromone units for different values of $k$. Parameters: $\gamma=0.003, s_{th}\in (16,40), t_{att}=0, \Delta t_v=10$.}\label{FIG policy kill rate}
\end{figure}

When confronted with predators that are less harmful for the colony (modeled with $k=0.5$, as opposed to our default value of $k=1$), the probability of stinging is lower, with a stronger difference at high pheromone concentrations (see Fig.~\ref{FIG policy kill rate}), since a more conservative reaction is preferable to avoid over-stinging. These lower probabilities lead to a slower response, which pays off in this case where the predator kills less bees per time step. Thus, we conclude that strong reactions to the alarm pheromone are developed when a defensive response needs to be built up as fast as possible. 

\subsubsection{Diversity of predators}\label{SUBSEC diversity of predators}
\begin{figure}[htb!]
\includegraphics[width=3.4in]{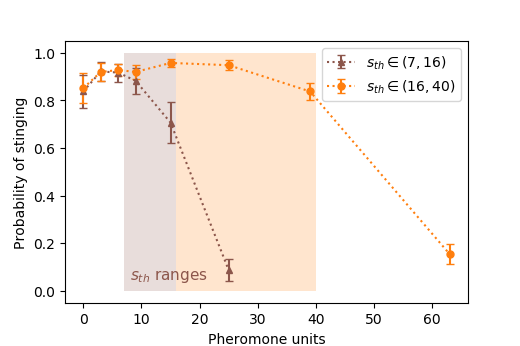}
\caption{Comparison of the learned probabilities of stinging for different ranges of $s_{th}$. Parameters: $\gamma=0.003, k=1, t_{att}=0, \Delta t_v=10$.}\label{FIG policy comparison sth}
\end{figure}

Honeybee colonies attract a wide array of vertebrate predators, from mice and toads to humans and bears. Obviously some are easier to deter than others, and their distribution may vary depending on the ecosystem considered. We evaluate the influence of predator diversity in different habitats by changing the range of $s_{th}$. In particular, we consider two learning processes, with ranges $s_{th}\in (7,16)$ and $s_{th}\in (16,40)$, which are chosen to correspond to percept bins (see Table~\ref{TAB:logbinning}). The former and the latter cases would represent habitats in which colonies face only small/easy to deter or only large/resistant predators, respectively. The resulting probabilities at the end of the learning processes are shown in Fig.~\ref{FIG policy comparison sth}, and they are very different. It appears that each population adapted primarily to the most resistant predator encountered, with bees stinging with very high probability until they reach an alarm pheromone concentration at which all predators should be gone. While it makes sense that honeybee colonies need to be able to cope with the worst predator in their environment, it is interesting to note that "weak" predators have little impact on the defensive strategy adopted.

\begin{figure}[htb!]
\includegraphics[width=3.4in]{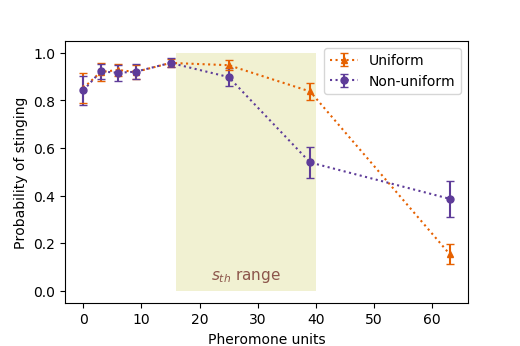}
\caption{Comparison of the learned probabilities of stinging when $s_{th}$ is uniformly distributed over the range $(16,40)$ and when the values $s_{th}\in (16,26)$ are four times more probable than $s_{th}\in (27,40)$. Parameters: $\gamma=0.003, k=1, t_{att}=0, \Delta t_v=10$.}\label{FIG policy nonuniform}
\end{figure}

\begin{figure}[htb!]
\includegraphics[width=3.4in]{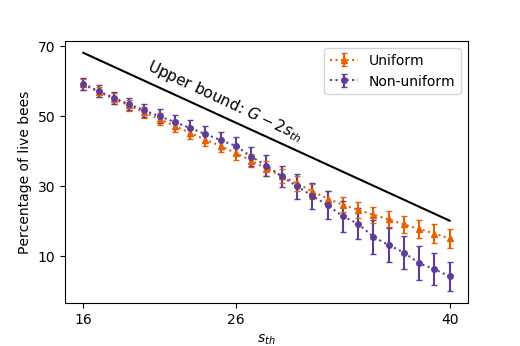}
\caption{Percentage of live bees at the end of two different learning processes where (i) $s_{th}$ is uniformly distributed over the range $(16,40)$ and (ii) the values $s_{th}\in (16,26)$ are four times more probable than $s_{th}\in (27,40)$. The upper bound indicates the optimal performance for each value $s_{th}$, which considers that exactly $s_{th}$ bees sting in the first $s_{th}$ time steps (i.e. $s_{th}$ bees are killed by the predator). Average and one standard deviation are obtained from data of the last 500 trials of the learning process. Parameters: $\gamma=0.003, k=1, t_{att}=0, \Delta t_v=10$.}\label{FIG live bees nonuniform}
\end{figure}

We also considered scenarios where certain types of predator appear more often than the rest. To model this, we consider the range $s_{th}\in (16,40)$ and two different learning processes: one where each $s_{th}$ in the range $(16,40)$ has the same probability to appear and another where $s_{th}\in (16,26)$ appears four times more often than $s_{th}\in (27,40)$. The results are given in Fig.~\ref{FIG policy nonuniform}, where we observe that the bees that encounter "weak" predators more often have a lower probability of stinging at high pheromone concentrations. More precisely, $p_s$ remains close to the initial value of $0.5$ for concentrations larger than 26 units, since they do not perceive such concentrations often. Fig.~\ref{FIG live bees nonuniform} shows how these colonies lose more members when attacked by the most resistant predators than those colonies that confronted them more often. Hence, the relative abundance of the different predators in a given environment also influences the defensive strategy adopted.

\subsubsection{Predation rate} \label{SUBSEC Predation rate}
\begin{figure}[htb!]
\includegraphics[width=3.4in]{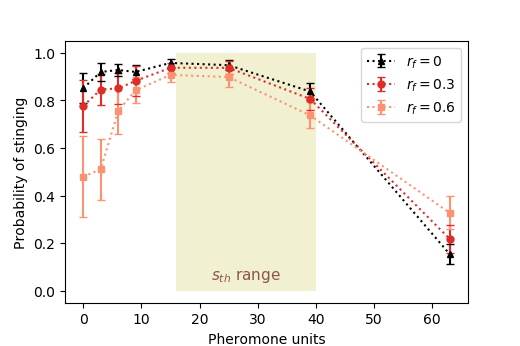}
\caption{Comparison of the learned probability of stinging for learning processes with different rates of false alarms. Parameters: $\gamma=0.003, s_{th}\in (16,40), k=1, t_{att}=0, \Delta t_v=10$.} \label{FIG policy false alarm}
\end{figure}

 In our model, each trial corresponds to a defensive event, which we assume is started by a non-specified disturbance close to the colony (for example, the visual perception of an object moving in the vicinity). In reality, most of these stimuli are likely to be unrelated to predation: they could be animals just passing by, falling branches and so on. Reacting to these stimuli would mean that bees waste time and efforts that could be better invested (and at worse that they may even die from stinging unnecessarily). The frequency of these \textit{false alarms} may depend on the environment considered: a high density of predators and/or the presence of specialized predators may be translated by a high predation rate for colonies, and hence fewer false alarms. To explore this, we include in our model trials in which there is no predator, which appear with probability $r_{f}$. Formally, for these trials $s_{th}$ is set to $0$, since there is no actual need to sting. We consider three learning processes, where the rate of false alarms $r_{f}$ is either $0, 0.3$ or $0.6$. The resulting behaviors are shown in Fig.~\ref{FIG policy false alarm}. We observe that the learned behavior for low pheromone concentrations changes drastically depending on the rate of false alarms. As the percentage of false alarms in the learning process grows larger, the initial probability of stinging gets lower to avoid over-reactions. However, the rise in stinging likelihood is then steeper, allowing the bees to quickly mobilize their nestmates in case of a real predator. 

\subsubsection{Defense termination}\label{SUBSEC defense termination}

Finally, we consider how bees determine when to \emph{stop} stinging. This can be explored by analyzing the effect of the parameter $\Delta t_v$ in our model. If $\Delta t_v=0$, then the agents visually perceive that the predator is leaving immediately after $s_{th}$ is reached. In this case, we expect that the bees always stop stinging after exactly $s_{th}$ stings, based on the "$v_{\text{ESC}}$" percept. Large values of $\Delta t_v$, on the other hand, represent the cases in which the bees do not have such a clear visual percept (for example, because the predator needs time to actually move away from the defended area). With $\Delta t_v=\infty$, the bees have no visual feedback about the success of their attack, hence their only way to survive is to sting enough that any predator would be deterred (maximum $s_{th}$ encountered). This is indeed what happens, and we provide the results of these two limiting cases ($\Delta t_v=0$ and $\Delta t_v=\infty$) in Appendix~\ref{APP extreme cases time delay}. In this section, we analyze the effects with intermediate values, $\Delta t_v=10,20$, which are likely closer to reality and where the optimal response is not obvious. 

\begin{figure}[htb!]
\includegraphics[width=3.4in]{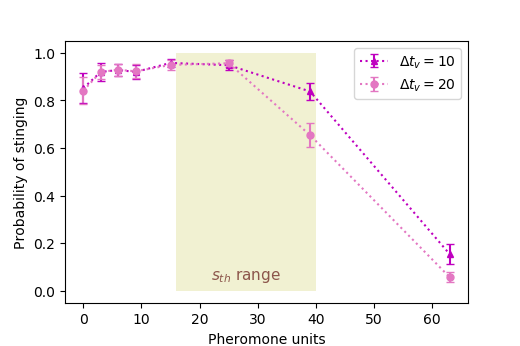}
\caption{Learned probability of stinging as a function of the pheromone units for different values of $\Delta t_v$. Parameters: $\gamma=0.003, s_{th}\in (16,40), k=1, t_{att}=0$.}\label{FIG policy comparison r}
\end{figure}

\begin{figure}[htb!]
\includegraphics[width=3.4in]{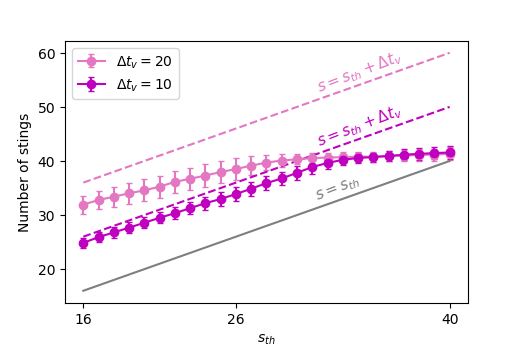}
\caption{Average number of stings depending on the predator size ($s_{th}$) for different values of $\Delta t_v$. Average and one standard deviation are obtained from data of the last 500 trials of the learning process. Dashed lines indicate the maximum number of times agents could have stung before the percept "$v_{\text{ESC}}$" is activated. Parameters: $\gamma=0.003, s_{th}\in (16,40), k=1, t_{att}=0$.}\label{FIG stings comparison r}
\end{figure}

Fig.~\ref{FIG policy comparison r} shows the probabilities at the end of the learning process. We observe how agents trained with $\Delta t_v=20$ have lower probabilities of stinging than those with $\Delta t_v=10$ for high pheromone concentrations (from 26 to 64 units). Our interpretation is that since it takes more time for these populations to realize that the predator is fleeing through the visual percept, they are at a higher risk of "wasting" bees (i.e. bees stinging ---and hence dying--- even though $s_{th}$ was already reached). To compensate, they start relying on high pheromone concentrations as a signal that an efficient defensive response has already been achieved. This self-limitation results in a number of stings close to the optimal performance for high values of $s_{th}$ (Fig.~\ref{FIG stings comparison r}). Thanks to this adaptation, the two populations perform similarly against resistant predators. 
On the other hand, decreasing the probability of stinging at low pheromone concentrations to avoid over-stinging the less resistant predators is not possible: if they did so, the agents would be less efficient against the most resistant predators. Therefore, they can only rely on the activation of "$v_{\text{ESC}}$" to stop stinging when dealing with weak predators. The direct consequence of this limitation is that, as $\Delta t_v$ increases, the bees over-sting more small/weak predators (Fig.~\ref{FIG stings comparison r}). The more pronounced decrease in stinging probability in populations with $\Delta t_v=20$ allows them to correct slightly this over-reaction, but they still waste more bees than populations with $\Delta t_v=10$. To conclude, the time needed for bees to perceive that the predator has moved away from their nest and is not threatening anymore is a limiting factor in their performance, but only when considering the weaker predators in a given environment.

\subsubsection{Size of defended area}
\begin{figure}[htb!]
\includegraphics[width=3.4in]{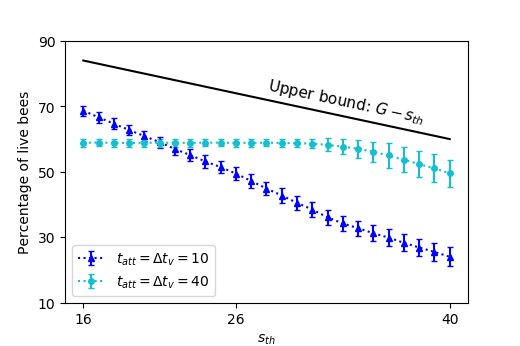}
\caption{Comparison of the percentage of live agents for different sizes of the defended territory. Colonies that defend a larger area (circles) perform better when confronted with resistant predators, but they over-sting the weak ones, which leads to a worse performance than colonies with smaller territories (triangles). Solid line indicates an upper bound for the performance of the colony, which considers that exactly $s_{th}$ bees sting and no bees are killed by the predator. Average and one standard deviation are obtained from data of the last 500 trials of the learning process. Parameters: $\gamma=0.003, s_{th}\in (16,40), k=1, t_{att}=\Delta t_v$.}\label{FIG comparison defended area}
\end{figure}

\begin{figure}[htb!]
\includegraphics[width=3.4in]{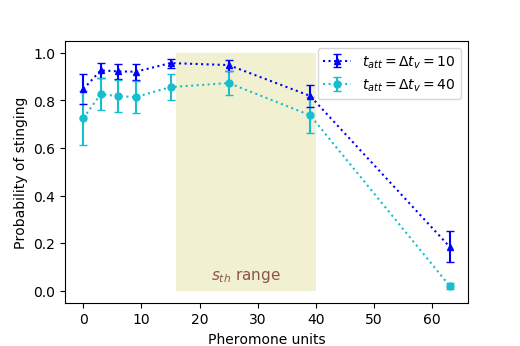}
\caption{Learned probability of stinging as a function of the pheromone units for different sizes of the defended area. Parameters: $\gamma=0.003, s_{th}\in (16,40), k=1$.}\label{FIG policy comparison defended area}
\end{figure}

Both $t_{att}$ and $\Delta t_v$ are linked to the time needed for the predator to move from the edge of the defended territory to the nest itself. Hence we could expect that for a given population, $t_{att}=\Delta t_v$ and that this single value represents the size of the defended area. It is interesting to note that an optimal behavior would require a high value for $t_{att}$ (to get a better chance to cope with predators with high $s_{th}$) but a low value for $\Delta t_v$ (to avoid over-stinging predators with small $s_{th}$). This is indeed what we observe in Fig.~\ref{FIG comparison defended area}, where colonies with $t_{att}=\Delta t_v=10$ (small area) perform better when confronted with less resistant predators but are worse at deterring predators with large $s_{th}$. Similarly to Fig.~\ref{FIG policy comparison t}, the probability of stinging is in general lower for the case with $t_{att}=\Delta t_v=40$ (see Fig.~\ref{FIG policy comparison defended area}). As a consequence of this analysis, we can postulate that defending a large area is an adaptation to predators that are on average quite resistant, whereas defending a small area is a better strategy when the range of predators encountered spans low values of $s_{th}$.

\subsection{Interpretation of the experimental results}
From the analyses performed in the previous section, we can compare the obtained responses with the experimentally observed ones (Fig.~\ref{FIG exp IAA response}) to study which conditions led to such behavior.

In general, we observe that most of the response curves in Sec.~\ref{SUBSEC predators parameters} (e.g. Figs.~\ref{FIG policy comparison sth} and \ref{FIG policy comparison r}) follow the same pattern as the experimental curve of Fig.~\ref{FIG exp IAA response}, that is, they start at a lower value and then increase until a maximum is reached (ramp-up phase), after which they decrease and get to a value that is non-zero. More precisely, we can analyze under which conditions these features emerge. A crucial observation can be made from Fig.~\ref{FIG policy false alarm}, where we show that the initial probability of stinging drops when the percentage of "false alarms" increases. This leads to a more pronounced increase of $p_s$ than in the simulations with no "false alarms". This can be interpreted as follows: since overreaction has a cost for the colony, the individual bee has a low probability of stinging when there is no alarm pheromone, which prevents many bees from leaving the hive after every minimal signal of danger, but the pronounced increase in $p_s$ enables a fast collective response when there is a predator to be driven away. Hence, the steepness of the ramp-up phase gives us insights about the predation rate to which populations were subjected in a given environment.

Finally, based on Fig.~\ref{FIG policy comparison sth}, we observe that $p_s$ starts decreasing at certain pheromone concentrations that vary depending on the total number of stings needed to scare away the largest or most resistant predators that the colonies face during the learning process. Thus, measuring the alarm pheromone concentration at which peak responsiveness is reached may provide information about the existence of specialized predators. In addition, from the interplay between the visual and the olfactory perceptions seen in Fig.~\ref{FIG policy comparison r}, we conclude that bees may self-limit their stinging responses once high pheromone concentrations are reached. If so, this likely depends on how much they rely on other perceptual information, such as visual stimuli, to time the end of defensive behavior. In particular, in the case where bees stop stinging based only on the visual input ($\Delta t_v=0$), we observe that the decay of the curve is simply a return to the baseline response (see Fig.~\ref{FIG extreme time delay} in App.~\ref{APP extreme cases time delay}).

\subsection{Case study: European vs. African bees}

\begin{figure}[htb!]
\includegraphics[width=3in]{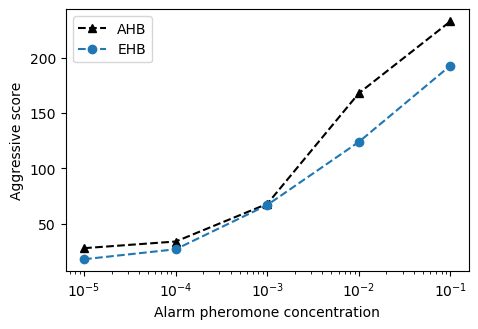}
\caption{Comparison of the aggressive score ($A_s$ ---see Sec.~\ref{SUBSEC AHBvsEHB}---) as a function of the alarm pheromone concentration for Africanized (AHB) and European (EHB) honeybees. Data modified from \citep{Collins87}. }\label{FIG exp responses AFRIvsEURO}
\end{figure}

Western honeybees (\textit{Apis mellifera}) split from other lineages 6 to 9 million years ago and spread into Africa and Europe, where they diversified \citep{Han12}. Among the many subspecies, \textit{A. m. scutellata} (that we call here "African bees" for simplicity) are well known for their fierce attacks. This species was introduced to the American continent, where they hybridized with bees previously imported from Europe. These "Africanized" bees, which retained the defensive behaviour of their African ascendants, have been widely studied because of the accidental deaths they caused, earning them the nickname of killer bees. While a few dozen bees may be recruited from so-called "European" colonies (e.g. \textit{A. m. carnica} or \textit{A. m. ligustica}), the same stimulus elicits stinging reactions from several hundreds Africanized bees. The latter also defend a much larger territory around their nest, and may chase the intruder for hundreds of meters (vs a few meters for European bees) \citep{Breed04, Guzman04, Collins82}. Importantly for our study, it was shown that African bees are more sensitive to SAP \citep{Collins87}. We slightly transformed the data from this previous study (see ~\ref{SUBSEC AHBvsEHB}) and plotted it in Fig.~\ref{FIG exp responses AFRIvsEURO} to allow easier comparisons with our results.

In this last section, we use our model to check several hypotheses about the evolution of these different defensive strategies. It has been hypothesized that several traits of African bees, including high defensiveness, evolved in response to higher predation rates in the tropics \citep{Winston92, Roubik89}. However this remains speculative: we found little information about the frequency of vertebrate predation on managed, present-day beehives, let alone assessments made on wild colonies. To assess the role of this evolutionary pressure, we set a higher predation rate for African bees in our model, or in other words a lower frequency of false alarms $r_f$.
Among their vertebrate predators, ratels (\textit{Mellivora capensis}, also called honey badgers) are often reported as one of the main threats to African beehives \citep{Carter07, Gebratsadik16, Schmidt14}. Because of their high resistance to bee stings, they are very successful at raiding colonies and account for a significant number of colony losses \citep{Carter07}. In temperate zones, bears also consume bees and can cause significant damage to beehives \citep{Clevenger92, Obrien90}. It is not clear how the impacts of these two predators compare. But the main predators of honeybee colonies are undoubtedly humans \citep{Schmidt14}. There, it is important to note that bees adapted to temperate climate (in Europe) establish large, perennial colonies. By contrast, African bees readily abandon their nesting site after a disturbance (a behavior termed absconding), making them hard to manage. As a consequence, while humans developed tools for the long-term keeping of bees in Europe, honey harvesting in Africa still largely results in the destruction of bee colonies \citep{Rinderer88}. Hence, humans and their protective gear can be considered as a highly resistant predator species in the African context, but as a mutualistic species in the European context. Several other primates are also among the toughest predators of African bees \citep{Schmidt14}. Finally, both subspecies regularly have to face smaller, less destructive predators such as mice, birds, badgers and reptiles. Taken together, this suggests that the predators of African bees are, on average, more difficult to deter than those of European bees. In our model, we translated this into a higher range of $s_{th}$ for African bees than for European bees.

In order to test whether these different properties can account for the differences in behavior observed between European and African bees, we modeled two types of bee populations that differed only with respect to these properties: frequency of false alarms ($r_f$) and range of $s_{th}$. We choose the populations to be more numerous ($G=200$) so that we can adopt different ranges of $s_{th}$ for the European and the African populations without reaching predator sizes that are too large for the group to drive away ($s_{th}>G/2$ when $k=1$ and $t_{att}=0$). We set $s_{th}\in (15,25)$ for the former and $s_{th}\in (15,70)$ for the latter to incorporate our assumption that African bees face more resistant predators. These ranges are chosen to match the edges of the corresponding percept bins, so that the performance is not limited by the sensory resolution. The probability of "false alarms" (see Sec.~\ref{SUBSEC Predation rate}) in the learning process for the European populations was set to $r_f^E=0.6$, against $r_f^A=0.3$ for the African ones. 

\begin{figure}[htb!]
\includegraphics[width=3.4in]{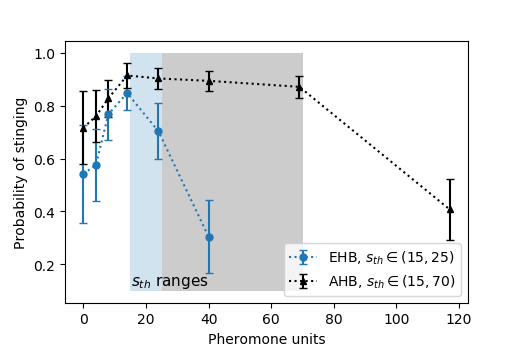}
\caption{Comparison of the learned probabilities of stinging between European and African populations. Shaded areas indicate the range of predators ($s_{th}$) that European (blue) and African (grey) colonies faced during the learning process. Average and standard deviation are obtained by taking data over 50 independently trained populations of 200 agents at the end of a learning process with $10^5$ trials. For clarity, percepts for which the probability of stinging remains at the initialization values ($p_s=0.5$) are not shown. Visual percept "$v_{\text{ESC}}$": $p_s=0.09\pm 0.01$ for EHB and $p_s=0.05\pm 0.01$ for AHB. Parameters: $\gamma=0.003, k=1, t_{att}=0, \Delta t_v=10, r_f^E=0.6, r_f^A=0.3$.}\label{FIG policy eurovsafri}
\end{figure}

The responses to alarm pheromone concentrations after the two learning processes are shown in Fig.~\ref{FIG policy eurovsafri}. We observe that African populations, which have learned to defend against larger, more frequent predators, develop a stronger reaction to the pheromone than European ones. The higher frequency of attacks in the African case makes the agents learn to react more strongly at low concentrations, which matches the experimentally observed responses (see Fig.~\ref{FIG exp responses AFRIvsEURO}). In addition, the probability of stinging remains high over a much larger range of alarm pheromone concentrations due to the larger $s_{th}$ interval encountered by African populations (see also Sec.~\ref{SUBSEC diversity of predators} and Fig.~\ref{FIG policy comparison sth}). 

\begin{figure}[htb!]
\includegraphics[width=3.4in]{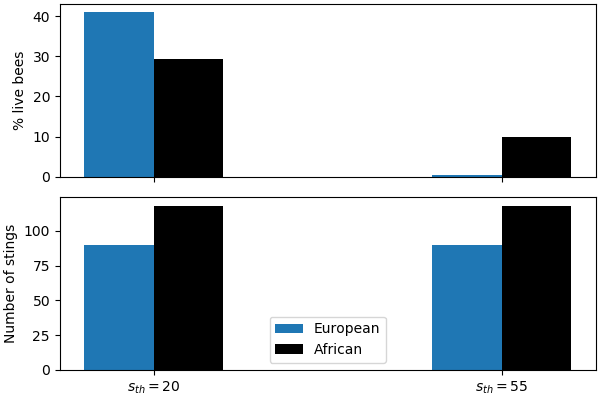}
\caption{Performance of European and African populations when tested after the learning process with predators with $s_{th}=20,55$. Note that European colonies have not confronted predators with $s_{th}=55$ during their learning process. Average is taken over 100 trials per population (50 independently trained populations are considered). For this test, we only consider the responses to the SAP, so $v_{\text{ESC}}$ is never perceived.}\label{FIG tested eurovsafri}
\end{figure}

To better understand the consequences of these different patterns of responsiveness to the alarm pheromone, we simulated ---once the respective learning processes are finished--- the collective responses of each population against two predators: one relatively easy to deter ($s_{th}=20$) and one highly resistant ($s_{th}=55$) (Fig.~\ref{FIG tested eurovsafri}). For this test, we consider the percept "$v_{\text{ESC}}$" to never get activated, so that we can analyze how the different responses to pheromone ---without the influence of other information--- affect the performance. We observe that the stronger reaction to pheromone of the African populations makes them overreact in the case with $s_{th}=20$, compared to the European bees. This is in line with the situation currently observed in the USA, where hybrids of these bees have been introduced: unprovoked attacks against humans or livestock are regularly reported, earning them the nickname "killer bees". On the other hand, they are able to survive the attack of the predator with $s_{th}=55$, whereas the European colony is eradicated. This suggests that the gentler European bees would not fare well in Africa, although to our knowledge such an experiment has not been attempted.

The above results show that our model is able not only to reproduce the experimental curve of Fig.~\ref{FIG exp IAA response}, but also to account for other types of behavior observed in different subspecies of the honeybee. These results are consistent with the experimental observation \citep{Collins87} that the African bees have stronger reactions to the alarm pheromone than the European bees and they support the hypothesis that the difference in aggressiveness at low pheromone concentrations is due to the higher attack rates of predators in the habitats of the African bees. Furthermore, under the assumption that African bees also encounter more resistant predators, we observe that they keep stinging over a broader range of alarm pheromone concentrations and that the decrease in their aggressiveness only starts at higher concentrations than for the European bees. It would be interesting to test African(ized) bees at these higher pheromone concentrations, like we did for European bees, in order to verify this prediction.

\section{Conclusions}

Social insects colonies are often called "superorganisms". This denomination refers to the fact that individuals within the colony coordinate their actions to reach fitness goals set at the colony level, effectively functioning as a single evolutionary unit. Understanding how this coordination is achieved and how group-level selection shaped the behavioral responses of individual group members is a fascinating and complex question. Here, we contribute to this field of research by combining experimental work and a novel computational approach to better comprehend the collective defensive behavior of honeybees.
In particular, we focused on responsiveness to the sting alarm pheromone, as this signaling mechanism is at the core of the bees' communication during a defensive event. First, we show experimentally that the stinging likelihood of individual bees varies depending on the concentration of SAP in the atmosphere. This response pattern exhibits at least two phases: an initial, ramp-up phase from low to intermediate concentrations of pheromone, followed by a decrease at high concentrations.
To interpret these results, we built a relatively simple agent-based model of the honeybee defensive behavior. The novelty of our approach resides in adapting Projective Simulation to a group of agents with a common goal (and hence a common reward scheme). We also added the constraint that all agents inherit the same decision process, as this better represents the heritability of aggressive traits across generations. This agent model allowed us to explore the impact of different evolutionary pressures on individual responsiveness to the alarm pheromone. From these insights, we postulate that the existence of the first phase (ramp-up) in SAP responsiveness results from a trade-off between avoiding false alarms and quickly recruiting nestmates in the presence of real predators. On the other hand, the decrease in stinging likelihood at high SAP concentrations could be due to a self-limiting mechanism to avoid unnecessary stings, or simply the consequence of a return to baseline because such high concentrations are never encountered in the wild, and hence no specific response had the chance to evolve. Finally, we also found that the SAP intensity at which the stinging probability peaks depends on the most resistant predator in a given environment.
Of course, our model is by no means exhaustive: there may be other environmental factors influencing the honeybees' defensive behavior that we did not investigate. Nonetheless, our study raises the interesting possibility that we could gain insights about the predation pressures suffered by specific populations of bees by measuring their dose-response curve to the alarm pheromone. We provide an example of this using previously collected data comparing African(ized) and European bees, and found that our model accurately predicted the experimental data when considering that African bees were subjected to higher predation rates. Experiments have also shown that African bees defend a much larger area around their nest. According to our model, this strategy is better suited to tackle resistant predators, again confirming that this subspecies probably evolved under strong predation constraints. Future studies could test this hypothesis by experimentally measuring the SAP concentration at which individual stinging responsiveness is maximum, and compare it to our data on European bees.
Altogether, our work provides new insights into the defensive behaviour of honeybees, and establishes PS as a promising tool to explore how selection on a collective outcome drives the evolution of individual responses. It opens the door for further investigations, both from the biological side and from the computational side. In particular, our model could be complexified to accommodate a few different types of individuals, thereby representing the different castes found in honeybee colonies.

\section{Acknowledgements}

MN was supported by a DFG research grant (project number 414260764) and by the Zukunftskolleg (University of Konstanz). The experimental data was gathered with the help of Maxime Pocher, Karoline Weich and Shehide Gashi. AL and HJB were supported by the Austrian Science Fund (FWF) through grant SFB BeyondC F7102. HJB was supported by the Ministerium für Wissenschaft, Forschung, und Kunst Baden-Württemberg (Az:33-7533-30-10/41/1). HJB and TM were supported by Volkswagen Foundation (Az:97721) and a BlueSky grant of the University of Konstanz.

\newpage
\bibliography{collective_sting_BIB}

\appendix

\section{Limiting cases with time delay $\Delta t_v=0,\infty$}\label{APP extreme cases time delay}

\begin{figure}[htb!]
\centering
\subfigure[]{\includegraphics[width=3in]{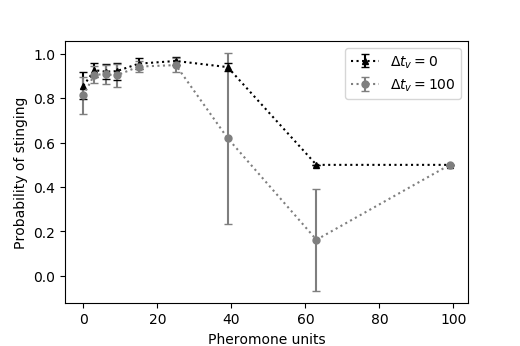}}
\subfigure[]{\includegraphics[width=3in]{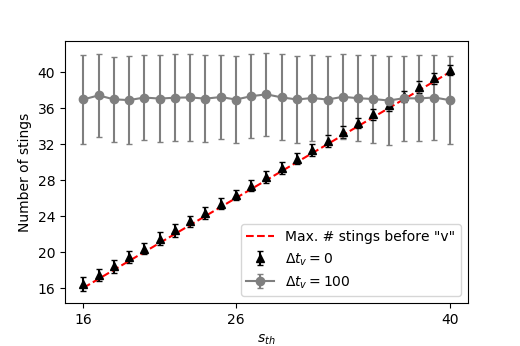}}
\caption{Comparison of the limiting cases where $\Delta t_v=0,100$. (a) Learned probabilities as a function of the pheromone units. (b) Number of stings as a function of predator sizes. Dashed, red line indicates the maximum number of times agents can sting before percept "$v_{\text{ESC}}$" is activated. Averages are taken over 50 independently trained populations of 100 agents. Parameters: $\gamma=0.003, s_{th}\in (16,40), k=1, t_{att}=0$. }\label{FIG extreme time delay}
\end{figure}

In this section, we provide the results of the learning processes with $\Delta t_v=0,\infty$, which are the limiting cases where the agents perceive the visual percept "$v_{\text{ESC}}$" immediately after $s_{th}$ is reached ($\Delta t_v=0$) and the one where "$v_{\text{ESC}}$" is never activated ($\Delta t_v=\infty$ ---technically, $\Delta t_v=100$ for this case with $G=100$---). Fig.~\ref{FIG extreme time delay} shows the learned probabilities as a function of the pheromone units and the average number of stings for each predator size. As expected, we observe how populations with $\Delta t_v=0$ learn to sting with high probability ($p_s\simeq 0.95$, see Fig.~\ref{FIG extreme time delay} (a)) for all the pheromone levels that they perceive (percepts 0 to 6) and they rely purely on the visual perception to stop stinging ---$p_s=0.005\pm 0.003$ for percept "$v_{\text{ESC}}$"---. We remark that these populations never encounter percepts 7 and 8 because percept "$v_{\text{ESC}}$" is immediately activated when $s_{th}$ is reached, and the largest predator they encounter has $s_{th}=40$. Therefore, the values of $p_s$ observed in Fig.~\ref{FIG extreme time delay} (a) for the last two points correspond to the initial ones, i.e. $p_s=0.5$, and the decay corresponds to a return to the preset probabilities and not to a self-limiting behavior. In Fig.~\ref{FIG extreme time delay} (b), we observe how these populations are able to sting the exact number of times $s_{th}$ for each predator size and they have no limitation based on the pheromone concentrations (triangular points lie on the dashed curve).

On the contrary, populations that have no visual cue of the predator leaving cannot adapt to different predator sizes and they learn to sting the number of times that is necessary to scare away the biggest predator they encounter (see Fig.~\ref{FIG extreme time delay} (b)). The limiting mechanism they develop is based only on the information about the pheromone concentration, which can be seen in Fig.~\ref{FIG extreme time delay} (a). There, we observe that there is a decay in the probability of stinging at percepts 6 and 7. The reason for the large standard deviations is that agents have no way of distinguishing the different predator sizes during the learning process, and they have to deal with contradictory information, since they will get lower rewards at trials with small $s_{th}$ if they sting too much, but they will all get killed if they do not reach the highest value of $s_{th}$ at trials with e.g. $s_{th}=40$.

\section{Cost of individual lives to the colony}
\begin{figure}[htb!]
\includegraphics[width=3.4in]{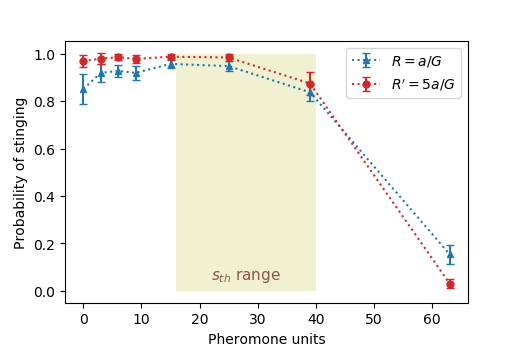}
\caption{Learned probability of stinging as a function of the pheromone units for different scalings of the reward function. Average and standard deviation are obtained by taking data over 50 independently trained populations of 100 agents. For clarity, percepts for which the probability of stinging remains at the initialization values ($p_s=0.5$) are not shown. Parameters: $\gamma=0.003, s_{th}\in (16,40), k=1, t_{att}=0, \Delta t_v=10$.}\label{FIG policy rescaled reward}
\end{figure}

Some subspecies of the honeybee have faster developmental rates and swarm more times per year \citep{Winston92}. Presumably, the cost of losing one bee for colonies of these subspecies is smaller than for other subspecies that do not reproduce as fast. In order to study the influence of the "life cost" on the defensive behavior, we re-scale the reward function of eq.~\eqref{EQ reward function} so that colonies that reproduce less get a higher reward per live bee. More precisely, we model more conservative colonies with $R'=5R$, i.e. the life of one bee is five times more valuable for this colony than for a colony with $R$. Fig.~\ref{FIG policy rescaled reward} shows how these bees develop a more extreme response to the alarm pheromone: they respond more strongly until the most resistant predator is scared away and they self-limit more once its $s_{th}$ is reached. The probability of stinging for percept "$v_{\text{ESC}}$" is also lower ($p_s=0.001\pm 0.001$) than the case with $R$ ($p_s=0.005\pm 0.002$). This behavior allows them to spare around 4\% more bees than the colonies with faster reproduction rates (i.e. with reward function $R$ in our model). 

\section{Additional data}\label{APP prob sting for v}

\begin{table}[htb]

\begin{tabular}{|c|c|c|}
\hline
$\gamma=0.001$ & $\gamma=0.003$ & $\gamma=0.007$ \\ \hline
$0.039\pm 0.007$ & $0.005\pm 0.002$ &  $0.012\pm 0.005$\\ \hline
\end{tabular}

\vspace{0.25cm}

\begin{tabular}{|c|c|c|}
\hline
9 percepts & 21 percepts & Full resolution \\ \hline
$0.005\pm 0.002$ & $0.006\pm 0.003$ &  $0.006\pm 0.002$\\ \hline
\end{tabular}

\caption{Probability of stinging for percept "$v_{\text{ESC}}$" for learning processes of Figs.~\ref{FIG policy calibration sth range} and \ref{FIG policy resolution}, respectively. See Sec.~\ref{SUBSEC learning process} in the main text for the rest of the parameters of each simulation.}\label{TABLE data ps for "v" (I).}

\end{table}

\begin{table}[htb]

\begin{tabular}{|c|c|}
\hline
$t_{att}=0$  & $t_{att}=60$ \\ \hline
$0.005\pm 0.002$ & $0.023\pm 0.011$ \\ \hline
\end{tabular}

\vspace{0.25cm}

\begin{tabular}{|c|c|}
\hline
$k=1$  & $k=0.5$ \\ \hline
$0.005\pm 0.002$ & $0.008\pm 0.005$ \\ \hline
\end{tabular}

\vspace{0.25cm}

\begin{tabular}{|c|c|c|}
\hline
$s_{th} \in (7,16)$  & $s_{th} \in (16,40)$ & $s_{th} \in (16,40) \text{  non-uniform}$\\ \hline
$0.018\pm 0.005$ & $0.005\pm 0.002$ & $0.007\pm 0.003$\\ \hline
\end{tabular}

\vspace{0.25cm}

\begin{tabular}{|c|c|c|}
\hline
$r_f=0$  & $r_f=0.3$ & $r_f=0.6$\\ \hline
$0.005\pm 0.002$ & $0.012\pm 0.006$ & $0.024\pm 0.010$\\ \hline
\end{tabular}

\vspace{0.25cm}

\begin{tabular}{|c|c|}
\hline
$\Delta t_v=10$  & $\Delta t_v=20$ \\ \hline
$0.005\pm 0.002$ & $0.006\pm 0.002$ \\ \hline
\end{tabular}

\vspace{0.25cm}

\begin{tabular}{|c|c|}
\hline
$\Delta t_v=t_{att}=10$  & $\Delta t_v=t_{att}=40$ \\ \hline
$0.007\pm 0.003$ & $0.024\pm 0.014$ \\ \hline
\end{tabular}

\caption{Probability of stinging for percept "$v_{\text{ESC}}$" for learning processes of Sec.~\ref{SUBSEC predators parameters}. See main text for details.}\label{TABLE data ps for "v" (II).}

\end{table}

\end{document}